\documentclass[11pt,a4paper]{article}
\usepackage{jinstpub}
\usepackage[export]{adjustbox}
\usepackage{makecell}
\usepackage{caption}
\usepackage{subcaption}

\usepackage{siunitx}
\DeclareSIUnit{\sqrtvolt}{\sqrt{\si{\volt}}}

\usepackage{miller}

\usepackage[linesnumbered,boxed]{algorithm2e}
\DontPrintSemicolon

\title{Results from a 2nd production run of low temperature wafer-wafer bonded pad-diodes for particle detection}

\author[1,2]{J.~W\"uthrich%
\note{Corresponding author.}%
\note{Now at the Physics Institute of the University of Z\"urich},}
\author{K.~Deplazes}
\author{and A.~Rubbia}

\affiliation{Institute for Particle Physics and Astrophysics, ETH Z\"urich,\\
Otto-Stern-Weg 5, Z\"urich, Switzerland}

\emailAdd{johannes.wuethrich@physik.uzh.ch}

\abstract{%
We are investigating the use of low temperature wafer-wafer bonding in the fabrication of next-generation particle pixel detectors.
This bonding technique could enable the integration of fully processed CMOS readout wafers with high-Z absorber materials, facilitating the creation of highly efficient X-ray imaging detectors.
It might also facilitate the integration of structures embedded inside the wafer bulk, such as deep uniform gain layers.
The bonding process results in a thin (nm-scale) amorphous layer at the bonding interface.
To study the impact of this interface on detector operation, we fabricated simple wafer-wafer bonded pad diodes using high resistivity float-zone silicon wafers.
Results from a first fabrication run of such diodes revealed that the presence of the bonding interface alters the depletion behaviour, with the interface acting as a heavily doped N++ layer.
However, metal contamination of the bonding surfaces during fabrication compromised these results, making them unrepresentative of an ideal bonding interface.
In this paper we present the results from a subsequent fabrication run, which does not exhibit this sort of metal contamination.
These results confirm that the bonding interface behaves as a heavily doped N++ layer, even without contamination.
Further, we discuss the reverse leakage current of the bonded samples, where current vs. voltage measurements show a behaviour which is consistent with current generated within the depletion region of the samples.
}


\begin{document}

\maketitle
\flushbottom

\section{Introduction}
\label{sec:introduction}

Low temperature wafer-wafer bonding (also called surface activated bonding in literature) is a technique which allows to directly fuse different semiconductor materials on a wafer basis at temperatures at or close to room temperature~\cite{takagi_surface_1996}.
As a type of direct bonding, no additional (non-conducting) material is present at the bonding interface.
The processing at room temperature allows to bond materials with different thermal expansion coefficients, as well as the bonding of fully processed CMOS wafers.
The successful integration of various material combinations has been shown extensively in literature~\cite{takagi_surface_1996,liang_annealing_2019,liang_electrical_2013,rebhan_200_2020,higurashi_room-temperature_2015}.

In the context of particle detection, especially X-ray imaging, this bonding technique is of interest, as it would allow to integrate absorber layers made of a high-Z material (such as GaAs) with a fully processed CMOS wafer containing the readout electronics~\cite{neves_towards_2019}.
Such a detector has a monolithic-like structure and charge collection, but is made of two different bulk materials.
It thus combines the advantages of a highly efficient absorber layer (high-Z) with an easy to fabricate readout chip (CMOS)~\cite{wuthrich_low-temperature_2023}.
Another potential application lies in the fabrication of special substrates with integrated uniform gain layers.

The monolithic-like integration of such detectors built with wafer-wafer bonding imposes that the bonding interface lies within the active volume, or between the active volume and readout node.
Thus it is of primordial importance to have sufficient understanding of how this interface influences properties such as signal collection, sensor depletion and leakage currents, in order to exploit this technology.
The nature of the bonding process imposes that a thin amorphous layer is formed at the bonding interface, created by the sputter cleaning of the bonding surfaces.
This amorphous layer usually has a thickness of a few \si{\nano\meter}.
The bonding process and the formation of this layer is described in more detail in~\cite{wuthrich_low-temperature_2023}.
Accordingly, the main challenge to the adoption of this bonding technique for the fabrication of particle detectors lies in gaining a proper understanding of the influence of this amorphous layer on the properties mentioned before.

\subsection{Initial fabrication run of wafer-wafer bonded pad diodes}
\label{sec:run2_summary}
In order to investigate the influence of the amorphous bonding interface, an initial fabrication run of wafer-wafer bonded pad diodes was carried out in the past.
This initial fabrication run is labelled Run~2 and the process and results were previously published in~\cite{wuthrich_low-temperature_2023}, \cite{wuthrich_depletion_2022} and \cite{wuthrich_tct_2023}.
The main results from these publications are summarized here.

\begin{figure}
  \begin{subfigure}[t]{0.65\textwidth}
    \includegraphics[width=\textwidth]{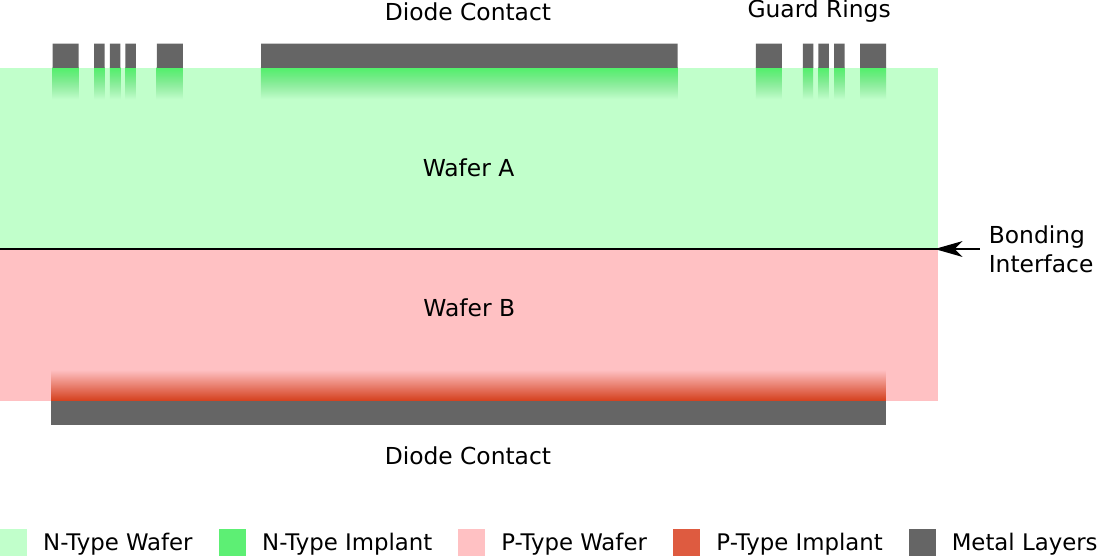}
    \caption{Schematic cross-section of the Run~2 bonded pad diodes.}
    \label{fig:initial_fabrication_xsection}
  \end{subfigure}
  ~
  \begin{subfigure}[t]{0.29\textwidth}
    \includegraphics[width=\textwidth]{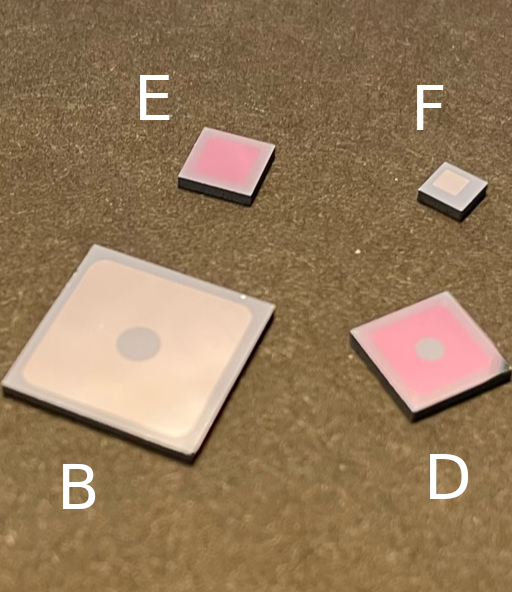}
    \caption{Fully processed diodes.}
    \label{fig:initial_fabrication_samples}
  \end{subfigure}

  \caption{Illustration of the fabricated diode samples (from~\cite{wuthrich_low-temperature_2023}).}
  \label{fig:initial_fabrication}
\end{figure}
The motivation to fabricate pad-diodes, compared to, for example, a fully integrated pixel detector, lies in the simplicity of the pad structures, allowing us to remove as many higher-order effects as possible and to allow for measurements of analog signals.
The pad diodes were fabricated by bonding high-resistivity float-zone silicon wafers of P-type to equivalent wafers of N-type.
The P-N junction used for the depletion of the sensor diodes is thus formed at the bonding interface.
To ensure good ohmic contacts, the P-side and N-side surface layers have patterned boron and phosphorus implantations, respectively.
This contact doping was created before bonding the two wafers; the metal contacts were formed via lift-off after the bonding process was completed.
A schematic cross-section of the fabricated samples is shown in figure~\ref{fig:initial_fabrication_xsection}.
Laser-dicing was used to separate the wafers into individual samples, which are shown in figure~\ref{fig:initial_fabrication_samples}.

The depletion behaviour of these samples was investigated with edge transient current technique (edge-TCT) measurements.
It was found that only the P-side of the bonded structure can be depleted.
This result indicates that the bonding interface acts as a highly doped N-type layer (N++) and thus inhibits the N-side of the bonded structure from depleting~\cite{wuthrich_tct_2023}.
Electron-dispersive X-ray spectroscopy measurements of the bonded samples showed that the bonding interface of the fabricated samples was contaminated with trace amounts of different metal elements (including iron, nickel and chromium).
Further investigations confirmed that these contaminants were introduced during the bonding process, due to the use of a wrongly configured bonding machine~\cite{wuthrich_tct_2023}.
During the bonding, the wafer surfaces to be bonded are cleaned via a low energy argon sputter beam.
In the bonding machine used, this beam also irradiated the stainless steel side walls of the vacuum vessel, leading to a re-deposition of the metal contaminants onto the bonding surfaces.
Generally, this problem can be mitigated by employing silicon coated shielding panels inside the bonding machine~\cite{jung_sige/si_2018}, but this was not the case in the machine used for Run~2.
The concentration of contaminants in the Run~2 samples is high enough that the possibility that the metal contaminants are responsible for the observed one-sided depletion could not be rejected~\cite{wuthrich_low-temperature_2023}.
Thus it was not clear if this N++ behaviour, and the resulting one-sided depletion, was an intrinsic property of the bonding interface, or solely due to the contamination.

Analysing the time-domain signal curves obtained via edge-TCT, a long exponential tail was observed on the recorded signals, which could not be explained with signal induction based on the Shockley-Ramo theorem~\cite{wuthrich_tct_2023}.
Further investigations showed that the presence of the undepleted (and thus non-isolating) N-side leads to this exponential tail.
An extension to the Shockley-Ramo theorem which takes into account conductive elements of the detector structure can be applied to this case~\cite{riegler_application_2019}.
In~\cite{wuthrich_low-temperature_2023} it was shown that a simple simulation model based on this extended Shockley-Ramo theorem fully reproduces the observed time-domain curves and is able to predict the signal from first principles by only accounting for the modified depletion behaviour.

In summary, the results from the initial fabrication run have shown that the presence of the bonding interface has a potential influence on the depletion behaviour of a detector.
The signals collected from such detectors can be explained from first principles when accounting for the modified depletion.
This indicates that the bonding interface does not seem to have any additional effect on the signal collection, beyond its influence on the depletion behaviour.

\section{Fabrication of samples without metal contamination}
\label{sec:run3_fabrication}
The discovery of the metal contamination in Run~2 led to an additional fabrication run (Run~3) whose results are presented in this paper.
The main goal is to reproduce the samples from Run~2, but without the metal contamination at the bonding interface.
This is achieved by reviewing the results from Run~2 and the bonding machine employed with the external company which carries out the actual bonding.
The types of wafers for this new production run are documented in table~\ref{tab:run3_wafers}.
\begin{table}
  \caption{Specifications of the wafers (silicon - \SI{100}{\milli\meter}) uses for the Run~3 production.}
  \label{tab:run3_wafers}
  \centering

  \begin{tabular}{lrrr}
    \textbf{Type} & \textbf{Thickness} (after polishing) & \textbf{Resistivity} & \textbf{Doping} \\ \hline
    P (Boron) \hkl<100> & \SI{503\pm3}{\micro\meter} & $> \SI{10}{\kilo\ohm\centi\meter}$ & $< \SI{2e12}{\per\centi\meter\cubed}$ \\
    N (Phosphorous) \hkl<100> & \SI{493\pm5}{\micro\meter} & $> \SI{10}{\kilo\ohm\centi\meter}$ & $< \SI{4e11}{\per\centi\meter\cubed}$
  \end{tabular}
\end{table}
The fabrication process of Run~3 follows the one of Run~2 as documented in~\cite{wuthrich_low-temperature_2023}.
In total four wafer pairs are part of Run~3.
The two pairs \textbf{P322422} and \textbf{P425325} have the same structure as the samples from Run~2, where the P-N junction is formed by bonding a P-type to an N-type wafer.
Wafer pair \textbf{P340329} is formed by bonding two P-type wafers, and the P-N junction is formed by using an N-type implant as the contact doping of one of the two wafers.
Similarly, wafer pair \textbf{P440429} is formed by bonding two N-type wafers, with one P-type contact implant to form the P-N junction.
All fabricated wafer pairs are summarized in table~\ref{tab:run3_wafer_pairs}.
\begin{table}
  \caption{Doping configurations of the wafer pairs fabricated for Run~3.}
  \label{tab:run3_wafer_pairs}
  \centering

  \begin{tabular}{c|rr|ll}
      & \textbf{Contact Implant} & \textbf{Wafer Bulk} & \textbf{Wafer Bulk} & \textbf{Contact Implant} \\ \hline
    \textbf{P322422} / \textbf{P425325} & P & P & N & N \\
    \textbf{P340329} & P & P & P & N \\
    \textbf{P440429} & N & N & N & P \\
  \end{tabular}
\end{table}
All samples are again diced with the same laser dicing process as used for Run~2~\cite{wuthrich_low-temperature_2023}.

\subsection{STEM and EDXS measurements}
\label{sec:run3_stem_edxs}
\begin{figure}
  \begin{subfigure}[t]{0.5\textwidth}
    \centering
    \includegraphics[width=\textwidth,clip,trim=0 0 0 15cm]{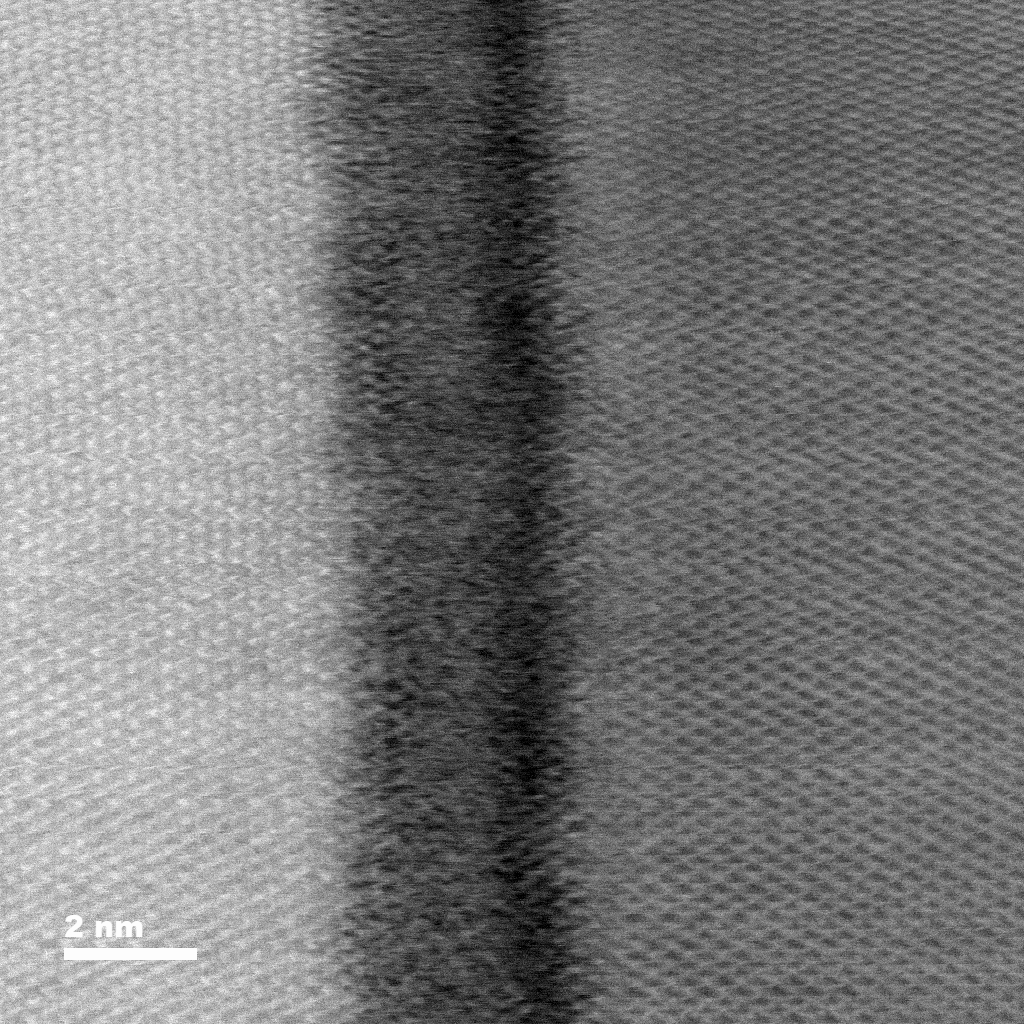}
    \caption{Run 2 - STEM}
    \label{fig:stem_exds_run2_stem}
  \end{subfigure}
  \begin{subfigure}[t]{0.5\textwidth}
    \centering
    \includegraphics[width=\textwidth,clip,trim=0 0 0 15cm]{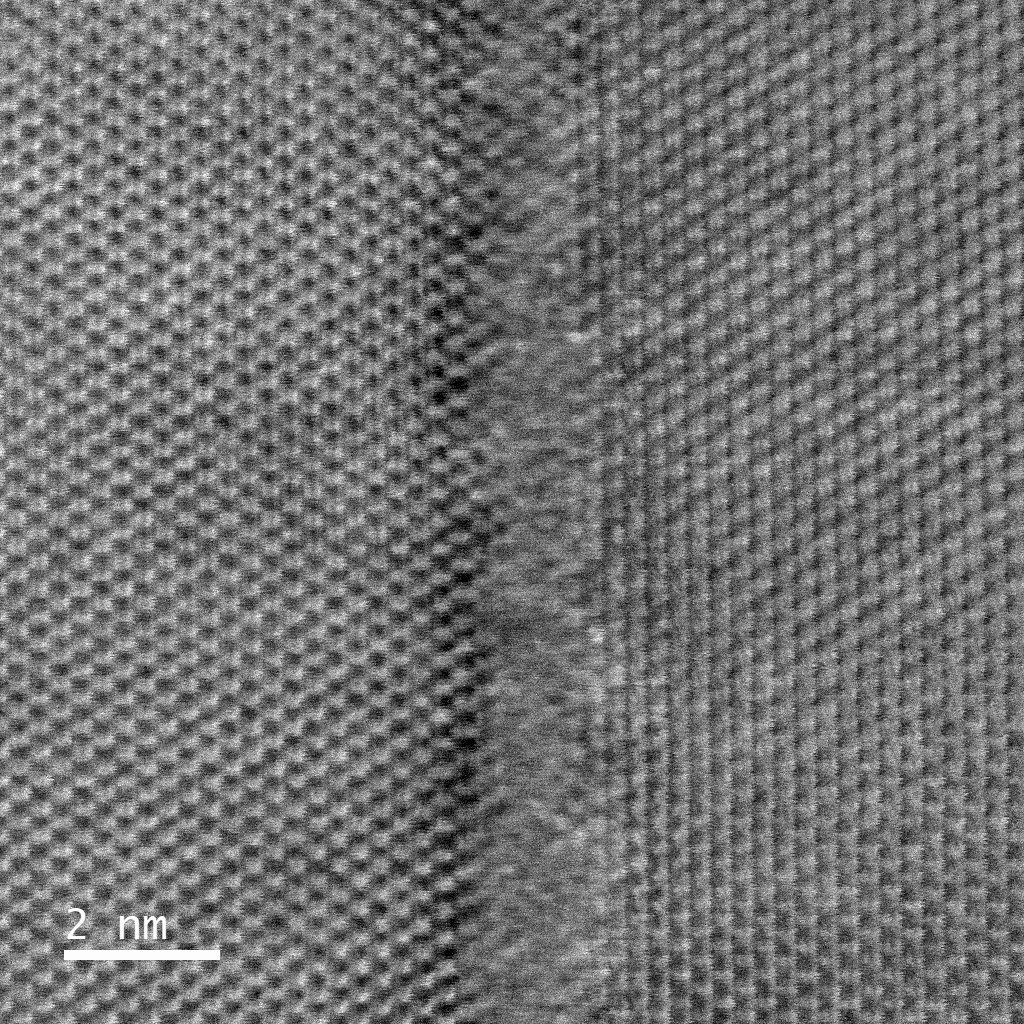}
    \caption{Run 3 - STEM}
    \label{fig:stem_exds_run3_stem}
  \end{subfigure}
  \\
  \begin{subfigure}[t]{0.5\textwidth}
    \centering
    \includegraphics[width=\textwidth,clip,trim=0 0 0 5cm]{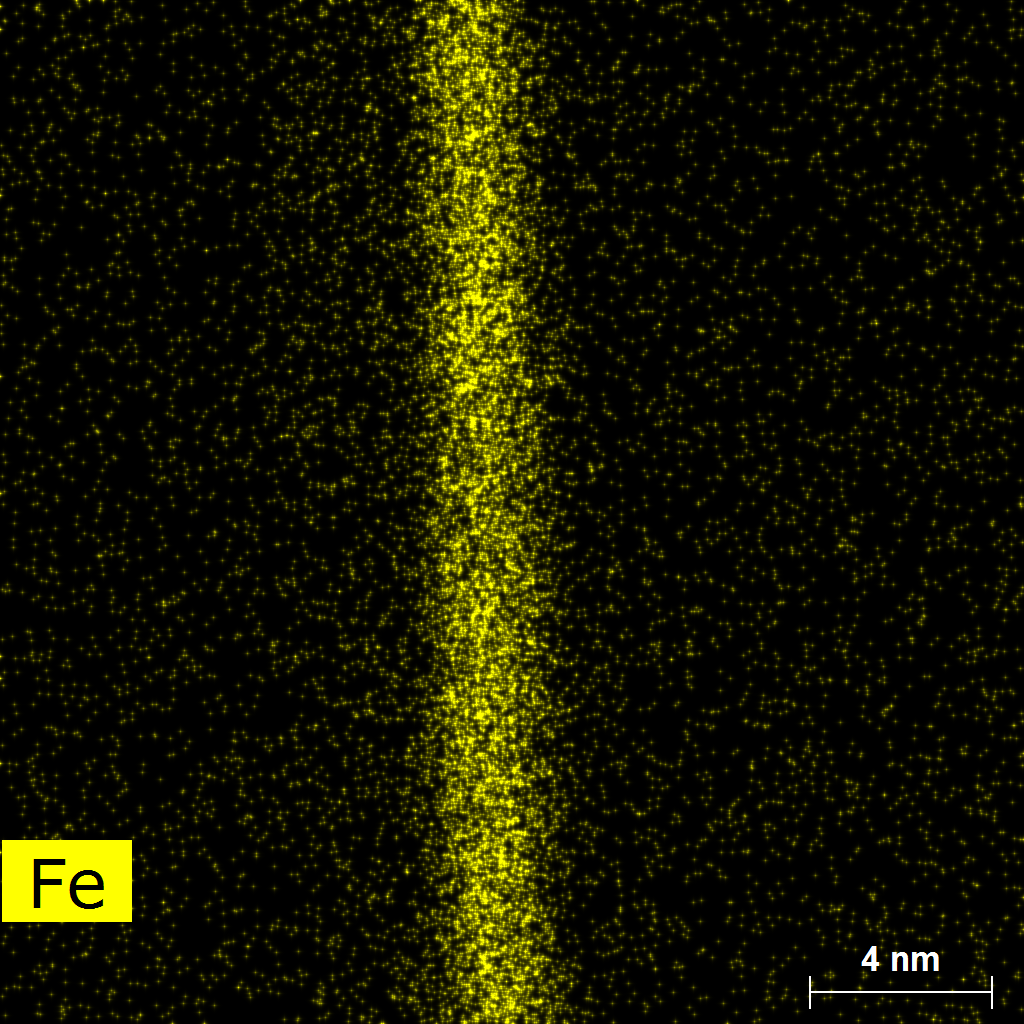}
    \caption{Run 2 - EDXS}
    \label{fig:stem_exds_run2_edxs}
  \end{subfigure}
  \begin{subfigure}[t]{0.5\textwidth}
    \centering
    \includegraphics[width=\textwidth,clip,trim=0 0 0 13.55cm]{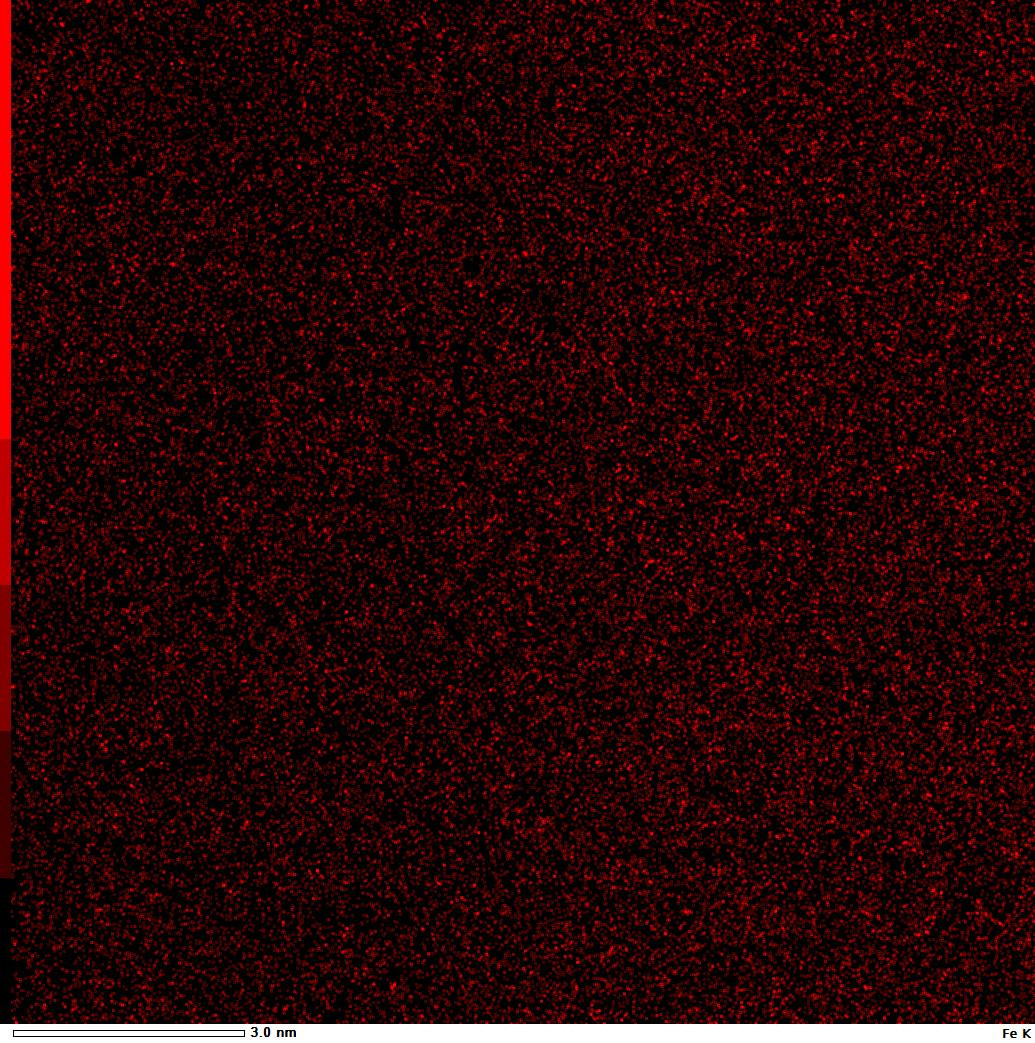}
    \caption{Run 3 - EDXS}
    \label{fig:stem_exds_run3_edxs}
  \end{subfigure}

  \caption{Bonding interface analysis using STEM and EDXS. The EDXS images show the distribution of iron. In Run~2 a clear contamination at the interface is visible. Run~3 does not show any contamination at the interface.}
  \label{fig:stem_edxs}
\end{figure}

As for the previous fabrication run, the quality and composition of the bonding interface were analysed using scanning transmission electron microscopy (STEM) and energy-dispersive X-ray spectroscopy (EDXS) respectively.
Samples from two different bonded wafers of Run~3 were analysed and showed the same results; one representative sample is shown here.
Figure~\ref{fig:stem_exds_run3_stem} shows the STEM image of the bonding interface of the new fabrication run.
Similarly to the results from Run~2 (shown in figure~\ref{fig:stem_exds_run2_stem}) a \SIrange{2}{3}{\nano\meter} thick amorphous layer is visible at the interface, which is expected for this type of bonding~\cite{neves_towards_2019}.
The EDXS results from the Run~3 samples showed no contaminating elements at the bonding interface.
The only element detected at the interface (in addition to Si) is Ar, a trace presence of which is expected as it is the element used for sputter cleaning the bonding surface during the bonding process.
A 2D composition map is shown in figure~\ref{fig:stem_exds_run3_edxs}, and a 2D map from Run~2 with metal contamination in figure~\ref{fig:stem_exds_run2_edxs}.
Thus, with Run~3 we successfully fabricated wafer-wafer bonded pad diodes without any detectable metal contamination.

\section{Measurement of the depletion behaviour with Edge-TCT}
\label{sec:tct_depletion}
\begin{figure}
  \includegraphics[width=\textwidth]{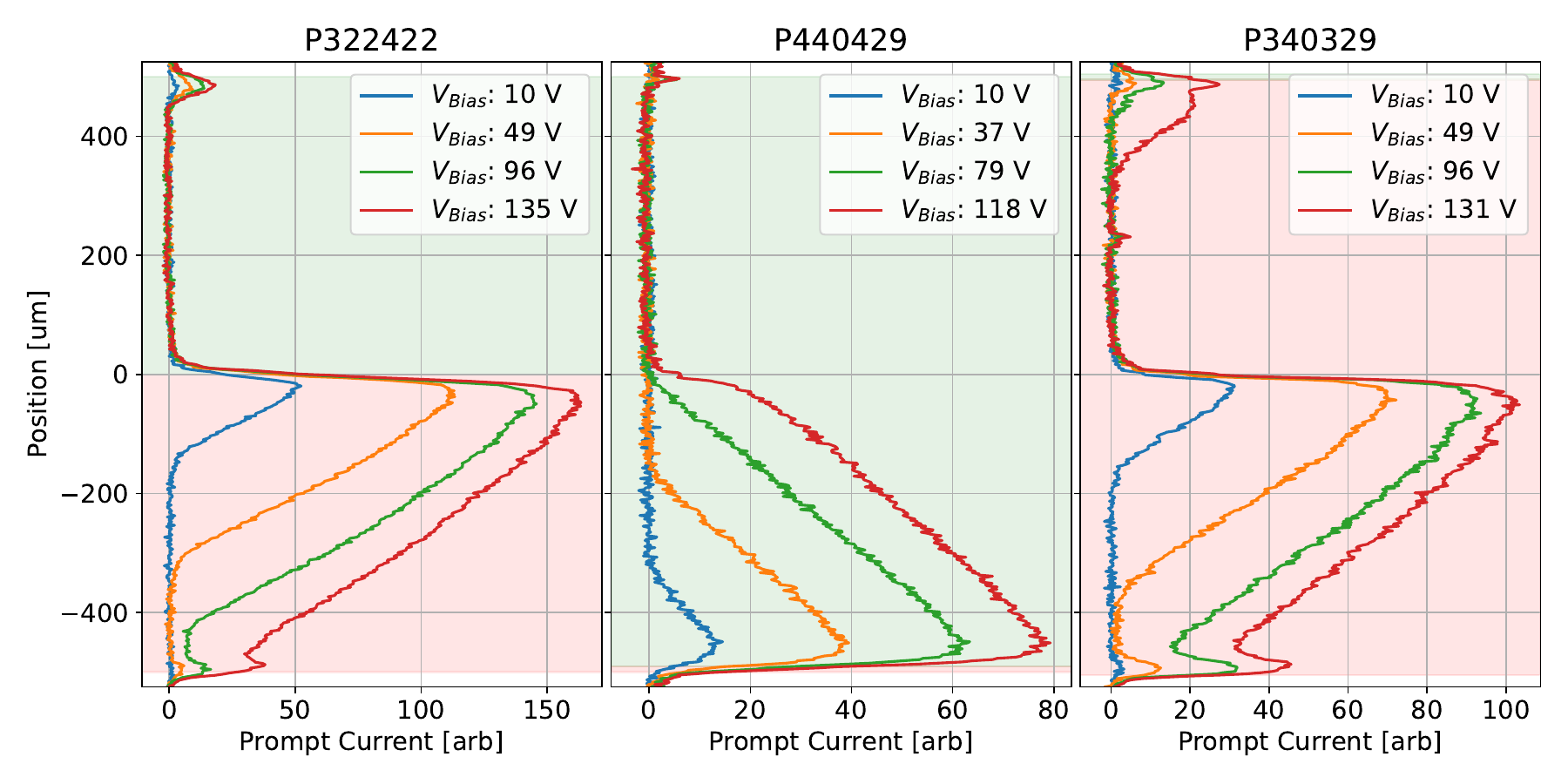}

  \caption{Prompt current curves for reverse biased samples of Run~3. Prompt current calculated according to~\cite{kramberger_investigation_2010} and~\cite{wuthrich_low-temperature_2023}.
  The prompt current values have an arbitrary scaling from sample to sample, as the effective prompt current depends on the number of generated electron/hole pairs as well as the substrate doping of a given sample.
  The data presented allows for a qualitative comparison of the depletion behaviour between the different samples.
  P-type regions shown with a red and N-type regions with a green background. The bonding interface is at $Y = \SI{0}{\micro\meter}$.}
  \label{fig:run3_tct}
\end{figure}

As for the samples of Run~2, the depletion behaviour of the Run~3 samples is investigated using edge-TCT.
For a detailed description of the TCT setup as well as the measurements procedure see~\cite{wuthrich_low-temperature_2023}.
Measurements of samples from the wafer pairs with a bonded P-N junction (\textbf{P322422} and \textbf{P425325}) show that again only the P-side of the structures can be depleted.
A total of eight samples taken from these two wafers were measured, all showing the same behaviour.
The prompt current curves of a representative sample are shown on the left of figure~\ref{fig:run3_tct}.
The depletion behaviour for the samples without metal contamination is thus the same as the previous samples, with the interface again exhibiting a N++ behaviour.

Edge-TCT measurements were also carried out with samples from the bonded wafer pairs with implanted P-N junctions.
The prompt current curves for a representative sample of \textbf{P440429} are shown in the center figure~\ref{fig:run3_tct}.
This sample is made of two bonded N-type wafers with a P-type contact implant.
As expected, the depletion region grows from the implanted P-N junction into the N-type bulk.
But the depletion region does not grow further once it reaches the bonding interface.
This is consistent again with the interface acting as a N++ layer, which would need to be fully depleted first, before the depletion grows beyond the interface.
For sample \textbf{P340329} the prompt current curves are shown on the right of figure~\ref{fig:run3_tct}.
Ignoring the presence of the bonding interface, the depletion is expected to start from the N-type contact implant at \SI{500}{\micro\meter}.
But the observed depletion again starts to grow at the bonding interface, with the depletion from the implanted P-N junction growing only when the opposite N-bulk is fully depleted.\footnotemark
\footnotetext{This sample can also be biased with opposite polarity. In which case the depletion region also starts from the bonding interface, but growing into the opposite side. This again is consistent with the N++ behaviour of the interface.}
This behaviour is again consistent with the interface acting as N++, leading to this sample having two stacked P-N junctions (one implanted and one at the interface).
In summary, all the measurements of the depletion behaviour of Run~3 samples confirm that the intrinsic interface acts as a strongly doped N++ layer.

\section{Reverse bias leakage current}
\label{sec:reverse_bias_current}

\begin{figure}
  \begin{subfigure}[t]{0.48\textwidth}
    \centering
    \includegraphics[width=\textwidth]{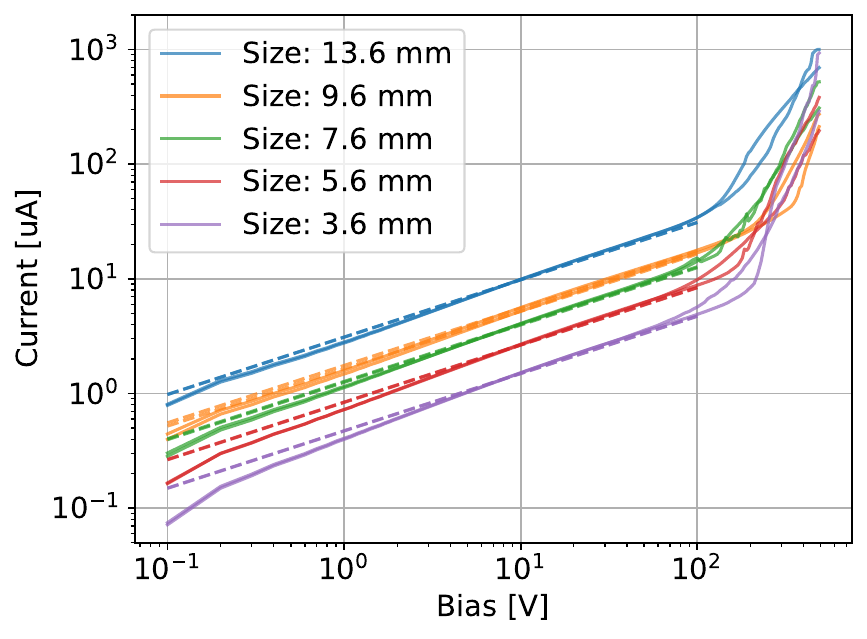}
    \caption{A subset of the measured I/V curves. At low voltages $V_{Bias} < \SI{100}{\volt}$ a clear separation is visible between diodes of different sizes. Dashes lines show the fitted functions $I_{rev} = A \sqrt{V_{Bias}}$.}
    \label{fig:run3_iv_examples}
  \end{subfigure}
  \hfill
  \begin{subfigure}[t]{0.48\textwidth}
    \centering
    \includegraphics[width=\textwidth]{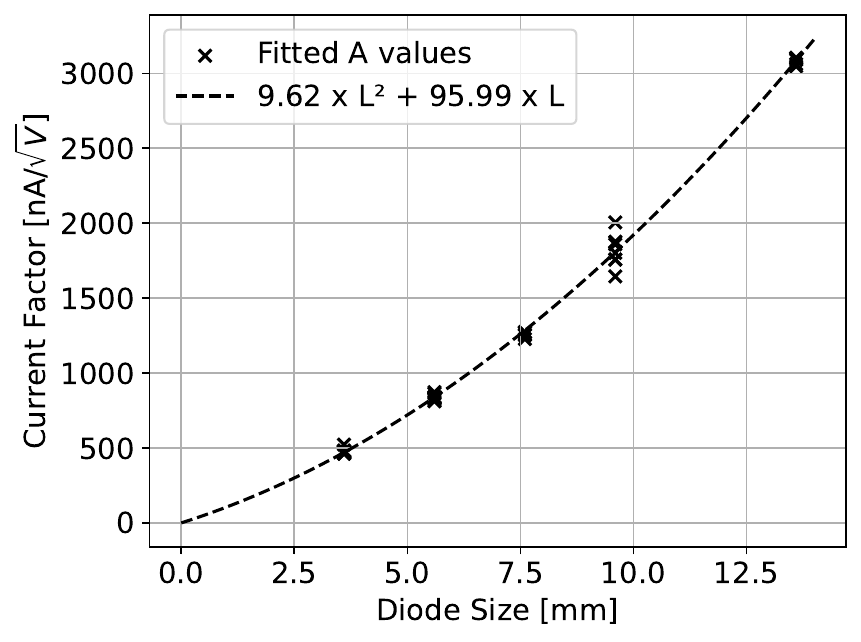}
    \caption{Fitted $A$ values in function of the diode size.
      The values show a clear quadratic dependency without offset, with $G_{bulk} = \SI{9.62}{\nano\ampere\per\milli\meter\squared\volt\tothe{-0.5}}$ and $G_{edge} = \SI{95.99}{\nano\ampere\per\milli\meter\volt\tothe{-0.5}}$.}
    \label{fig:run3_iv_fitted}
  \end{subfigure}

  \caption{Analysis of the reverse bias current of sample from \textbf{P322422}.}
  \label{fig:run3_iv}
\end{figure}

In~\cite{wuthrich_low-temperature_2023} it was shown that the bonded pad diodes indeed have rectifying properties with a forward bias current multiple orders of magnitude higher than the reverse bias current.
This also holds for the diodes fabricated for Run~3.
The reverse bias current for one Run~3 wafer with a bonded P-N junction (\textbf{P322422}) is studied here in more detail.
As documented in~\cite{wuthrich_low-temperature_2023} each wafer contains five different types of diodes, which have the same structure, but differ in size (edge length of \SIrange{3.6}{13.6}{\milli\meter}).
Of each diode type between 5 and 16 different samples were analysed,\footnotemark
\footnotetext{Fewer samples are available for the larger diodes due to the limited amount of space on a \SI{100}{\milli\meter} wafer.}
and some example curves are shown in figure~\ref{fig:run3_iv_examples}.
The measured data shows a clear $I_{rev} = A \sqrt{V_{Bias}}$ dependence at low bias voltages, this is indicated by the dashed fit lines in figure~\ref{fig:run3_iv_examples} and is a sign that the reverse leakage current at low voltages is primarily due to generation in the depleted bulk.
The fitted $A$ parameters for all available samples are plotted in function of the diode edge length $L$ in figure~\ref{fig:run3_iv_fitted}.
They show a clear quadratic dependency of the form $A(l) = G_{bulk} L^2 + G_{edge} L$ which is shown as the dashed line in figure~\ref{fig:run3_iv_fitted}, with the fitted $G_{bulk}$ and $G_{edge}$ parameters given in the legend.
Accordingly, we can conclude that the reverse bias current of the P-N bonded samples at low voltages follows the following model:
  $$I_{rev}\left(V_{Bias}\right) = \left(G_{bulk} L^2 + G_{edge} L\right) \sqrt{V_{Bias}} = g_{bulk} L^2 C_W \sqrt{V_{Bias}} + 4 g_{edge} L C_W \sqrt{V_{Bias}}$$
with the depletion width $W_{dep} = C_{W} \sqrt{V_{Bias}}$ and $C_W = \sqrt{\frac{2 \epsilon_{Si}}{q N_a}}$ for a one-sided abrupt P-N junction~\cite{sze_physics_2007}.
Thus, the observed reverse bias current is consistent with a model where the current is due to generation within the depleted bulk ($g_{bulk}$ term) and generation along the diode edges which are exposed to the depletion region ($g_{edge}$ term).

Temperature dependent measurements (in the range $T \in \SIrange{5}{40}{\celsius}$) of the I/V curve of one sample show that the reverse leakage current follows an exponential law in the form $I_{rev}(T) \propto \exp\left(-\frac{E_A}{k_B T}\right)$ with the activation energy $E_A \in \SIrange{0.5}{0.6}{\eV}$ for reverse bias voltages of $V_{Bias} \in \SIrange{0.1}{60}{\volt}$.
The observed activation energy $E_A$ is close to half the bandgap in silicon ($E_{gap} \approx \SI{1.1}{\eV}$) and thus the observed behaviour is consistent with the thermally excited generation current via deep level traps within the bandgap~\cite{chilingarov_generation_2013}.

Overall, the observed reverse bias current at low bias voltages shows a behavior that agrees well with current due to thermal generation within the depleted zone of the diodes.
Conversely, this indicates that the bonding interface only has a minor influence on the reverse bias current at these voltages.
On the other hand, the current observed at higher reverse bias voltages does not show any consistent behaviour, neither in function of the bias voltage, nor in function of the diode size (see figure~\ref{fig:run3_iv_examples}).
It is unclear if this inconsistent current behaviour at higher voltages is due to the bonding interface, or due to additional effects (including over the edges) once the depleted region reaches the surface of the sample (which occurs close to $\SI{100}{\volt}$).

\section{Conclusions}
\label{sec:conclusions}

We investigated the effect of the amorphous layer created during low temperature wafer-wafer bonding on the operating of a wafer-wafer bonded particle detector, namely on the depletion behaviour, signal formation and leakage current.
The results from an initial production run of wafer-wafer bonded pad diodes showed a modified depletion behaviour due to the presence of the bonding interface, but no direct influence of the bonding interface on signal formation.
As these initial samples exhibit trace metal contamination at the bonding interface these results could not directly be applied to the behaviour of a clean bonding interface.
A new production run was carried out to obtain samples without any metal contamination, which has been confirmed via EDXS measurements.
Edge-TCT measurements on these new samples show that the intrinsic (non-contaminated) bonding interface still leads to a modification of the depletion behaviour, with the interface acting as a highly doped N++ layer.
This observation is confirmed by multiple samples with either a bonded P-N junction or P-N junctions created via implantation.
We further observe that the reverse leakage current at low bias voltages shows a behaviour which is consistent with generation current in the bulk (and on the edges) of the samples.
This indicates that the bonding interface itself has little impact on the reverse leakage current at these low reverse bias voltages.
The reverse current at higher bias voltages does not show any consistent behaviour, and it is unclear if this is due to the presence of the bonding interface, or due the depletion zone reaching the opposite side contact.

\acknowledgments
The authors thank the Cleanroom Operations Team of the Binnig and Rohrer Nanotechnology Center (BRNC) for their help and support during the fabrication process.
We further thank Dr.~P.~Zeng and Dr.~E.~Barthazy of ScopeM for their support and assistance with the STEM and EDXS analysis of the samples.

\bibliographystyle{JHEP}
\bibliography{Run3Results_JWuethrich}

\end{document}